\documentclass{aa} 

\usepackage{natbib}
\bibpunct{(}{)}{;}{a}{}{,} 

\usepackage{stfloats}
\usepackage{array}
\usepackage{makecell}
\usepackage{placeins}
\usepackage{graphicx}
\usepackage{deluxetable} 
\usepackage{balance}
\usepackage{amsmath}
\usepackage[colorlinks=true,linkcolor=blue,citecolor=blue,urlcolor=blue]{hyperref}
\usepackage{txfonts}
\begin{document}

   \title{Sungrazer comets as analogs of star-planet magnetic interactions}
   
   \author{L-S. Guité
          \inst{1},
          A. Strugarek\inst{1}, A. J. Finley\inst{2}, S. Parenti\inst{1}, V. Réville\inst{3}, A. Paul\inst{1}, A. S. Brun\inst{1}, and J. de Carpentier\inst{4} 
          }

    \institute{Université Paris-Saclay, Université Paris Cité, CEA, CNRS, AIM, 91191, Gif-sur-Yvette, France \\
    \email{louis-simon.guite@cea.fr}
    \and
    European Space Agency, ESTEC, Noordwijk, The Netherlands
    \and
    Institut de Recherche en Astrophysique et Planétologie (IRAP), Université de Toulouse, CNRS, CNES, Toulouse, France
    \and
    Collège Stanislas, 75006 Paris, France
    }
    
\abstract{Star–planet magnetic interactions (SPMIs) can transfer energy from an exoplanet to its host star via Alfvén waves when the planet orbits within a sub-Alfvénic stellar wind. Similar conditions were encountered by the sungrazing comet Lovejoy as it passed through the solar corona in December 2011. The possibility that comets could trigger solar activity via magnetic interactions has never been investigated.}{The aim of this paper is to quantify the energetics of potential magnetic interactions between comet Lovejoy and the Sun, and to assess if enough energy could be deposited to either form a hotspot or trigger eruptions on the Sun.}{We used the magnetohydrodynamic WindPredict-AW model to reconstruct the coronal magnetic field and solar wind conditions along the comet’s orbit, and determined the magnetic footpoints that connected the comet to the solar surface. By estimating the travel time of hypothetical Alfvén waves, we identified a brightening event observed by EUVI aboard STEREO-A that is spatially and temporally consistent with comet Lovejoy’s passage. We then evaluated the associated energy budget using SPMI power scaling laws.}{We computed the SPMI power for all magnetic field lines whose footpoints are located within 5$^{\circ}$ of the brightening. The resulting power distribution spans approximately $10^{14}$–$10^{16}$ W. The power associated with the field line whose hypothetical Alfvén waves arrive just before the brightening lies at the lower end of this distribution. In comparison, the estimated radiative power of the brightening event in the STEREO-A/EUVI $195\AA$ channel is approximately $10^{17}$ W.}{We find that comet Lovejoy does not generate sufficient SPMI power to be the energy source of the observed brightening intensity. However, it may still act as a perturbation to existing magnetic structures, and lead to the triggering of solar flares. Confirming this hypothesis would require additional observations that do not exist for this event, which make future passages of sungrazing comets valuable opportunities to study SPMI processes within our solar system.}


\titlerunning{Sungrazer comets as analogs of star-planet magnetic interactions}
\authorrunning{Guité, L-S., et al.}

\maketitle

\section{Introduction}\label{section:introduction}

\par Star–planet magnetic interactions (SPMIs) have been both modeled \citep[see, for example,][]{Strugarek2018, Vidotto2025} and observed \citep[e.g.,][]{Shkolnik2008, Cauley2018, Cauley2019, Strugarek2019}, and offer a potential indirect means of probing exoplanetary magnetic fields. In the context of exoplanetary systems, these interactions occur when an exoplanet orbits within the Alfvén surface of its host star \citep[e.g.,][]{Zarka2007}, i.e., the region where the stellar wind velocity is sub-Alfvénic. In this case, Alfvén waves can be launched from the exoplanet and travel back toward the star along structures called Alfvén wings \citep{Neubauer1998, Strugarek2015}, if the exoplanet has a magnetosphere or is electrically conductive \citep{Laine2012, Strugarek2018}. Ultimately, these waves can dissipate and deposit energy in the stellar atmosphere \citep[e.g.,][]{Cuntz2000, Paul+2025}.
\par In the solar system, there are no sub-Alfvénic SPMIs as no planet orbits within the Alfvén surface of the Sun, which roughly lies between 10 to 20 solar radii \citep[e.g.,][]{Cranmer2023, Finley2025, Badman2025}. Moons orbiting Jupiter, such as Io or Ganymede, do exhibit bidirectional sub-Alfvénic energy transfer that leads to regions of enhanced emissions on the auroral oval of Jupiter \citep[e.g.,][]{Clarke1996, Cowley2001}, and this can be understood in the context of Alfvén wings \citep[e.g.,][]{Saur2013}. Recently, similar observations have been made for the moon Enceladus orbiting Saturn \citep{Kim2024,Hadid2026}. However, as shown by \cite{Paul2026}, these interactions are only rudimentary analogs to SPMIs, since the moons orbit within a planetary magnetosphere in the absence of solar wind, and under significantly smaller Alfvénic Mach numbers compared to those of exoplanetary systems. 

\par Comets can enter the Sun's Alfvén surface during their orbit. In the context of SPMIs, it is therefore reasonable to think that a comet with an ionized tail could perturb the local magnetic field and, as a consequence, launch Alfvén waves along connected field lines and deposit energy into the solar atmosphere. This energy could tentatively either form a localized hot spot \citep[e.g.,][]{Ip2004,Klein2022, Strugare2025rev} and/or destabilize a preexisting magnetic structure, potentially triggering an eruption \citep{Ilin2025}. Some comets, known as sungrazers, approach to within 2.45 solar radii of the solar surface (see \citealt{Jones2018} for a review on sungrazers), well inside the corona, which makes them prime candidates for detecting solar-system analogs of SPMIs.
\begin{figure*}[ht]
    \centering
    \includegraphics[width=1.3\columnwidth]{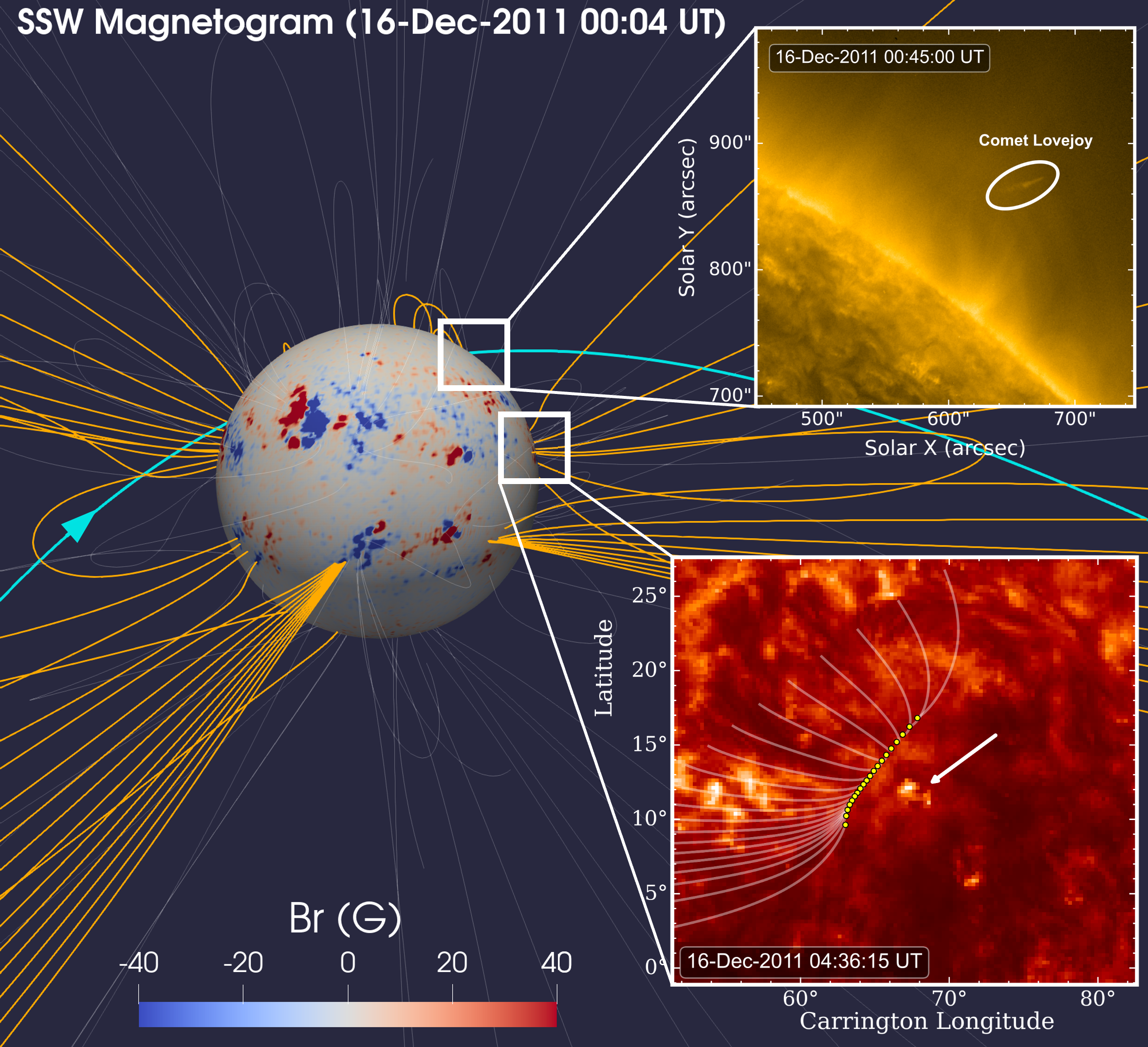}
    \caption{Configuration of the coronal magnetic field during the passage of comet Lovejoy, as seen from an Earth vantage point. The surface displays the photospheric radial magnetic field of a SolarSoft synchronic map, generated using a HMI magnetogram on December 16, 2011 and used to reconstruct the magnetic field using the WindPredict-AW MHD code. The orange field lines connect the comet to the solar surface, with its trajectory traced out by the cyan curve. The arrow on the trajectory indicates the direction of motion along the orbit. The zoomed-in section in the bottom right shows the brightening event of interest in this paper, marked by a white arrow, as observed by STEREO-A/EUVI in the $304\AA$ band, triggered on December 16, 2011. The field line footpoints are also shown as yellow dots. In the upper right, we see the cometary tail in an SDO/AIA $171\AA$ image during the comet's egress.}
    \label{fig:comet_trajectory}
\end{figure*}

\par To maximize the likelihood of such a detection, an ideal case would involve a comet that survives its perihelion passage, allowing for a prolonged interaction with the coronal magnetic field during a period of continuous space-based extreme ultraviolet (EUV) observations, which are essential for detecting any associated activity. Under these conditions, the 2011 perihelion passage of comet C/2011 W3 (Lovejoy) represents the most promising event. Several aspects of this event have been investigated, including the EUV emission \citep{McCauley2013, BryansPesnell2012}, UV spectroscopy \citep{Raymond2018}, and time-dependent chemistry of the plasma tail \citep{PesnellBryans2014}; the use of the tail morphology to probe the coronal magnetic field \citep{Downs2013, Raymond2014}; multi-fluid magnetohydrodynamic (MHD) modeling of the plasma tail \citep{Jia2014}; and analyses of the comet’s orbit and dust tails \citep{Sekanina2012}. However, the potential impact of magnetic interactions on solar activity during this event has not yet been studied.
\par In this paper, we investigate whether comet–Sun magnetic interactions could deposit energy in the solar corona in the form of a hotspot and/or trigger flaring activity. In Section~\ref{section:lovejoy}, we analyze the magnetic connectivity between comet Lovejoy and the Sun during its perihelion passage, for both potential field source surface \citep[PFSS;][]{Altschuler1969, Schatten1969} and MHD models, and estimate the energetics of potential magnetic interactions mediated by propagating Alfvén waves. In Section~\ref{section:discussion}, we situate our results in the broader context of SPMIs and identify future comets as potential candidates for such interactions. We summarize our findings in Section~\ref{section:conclusion}. In Appendix \ref{appendix_powers}, we examine the distribution of SPMI powers near the brightening event, while in Appendix \ref{appendix:orbit} we look at the solar wind Alfvénic Mach number and the SPMI power along the comet's orbit. Finally, we show the robustness of our results with respect to the chosen model parameters in Appendix \ref{appendix}.

\begin{figure*}[ht]
    \centering
    \includegraphics[width=1.7\columnwidth]{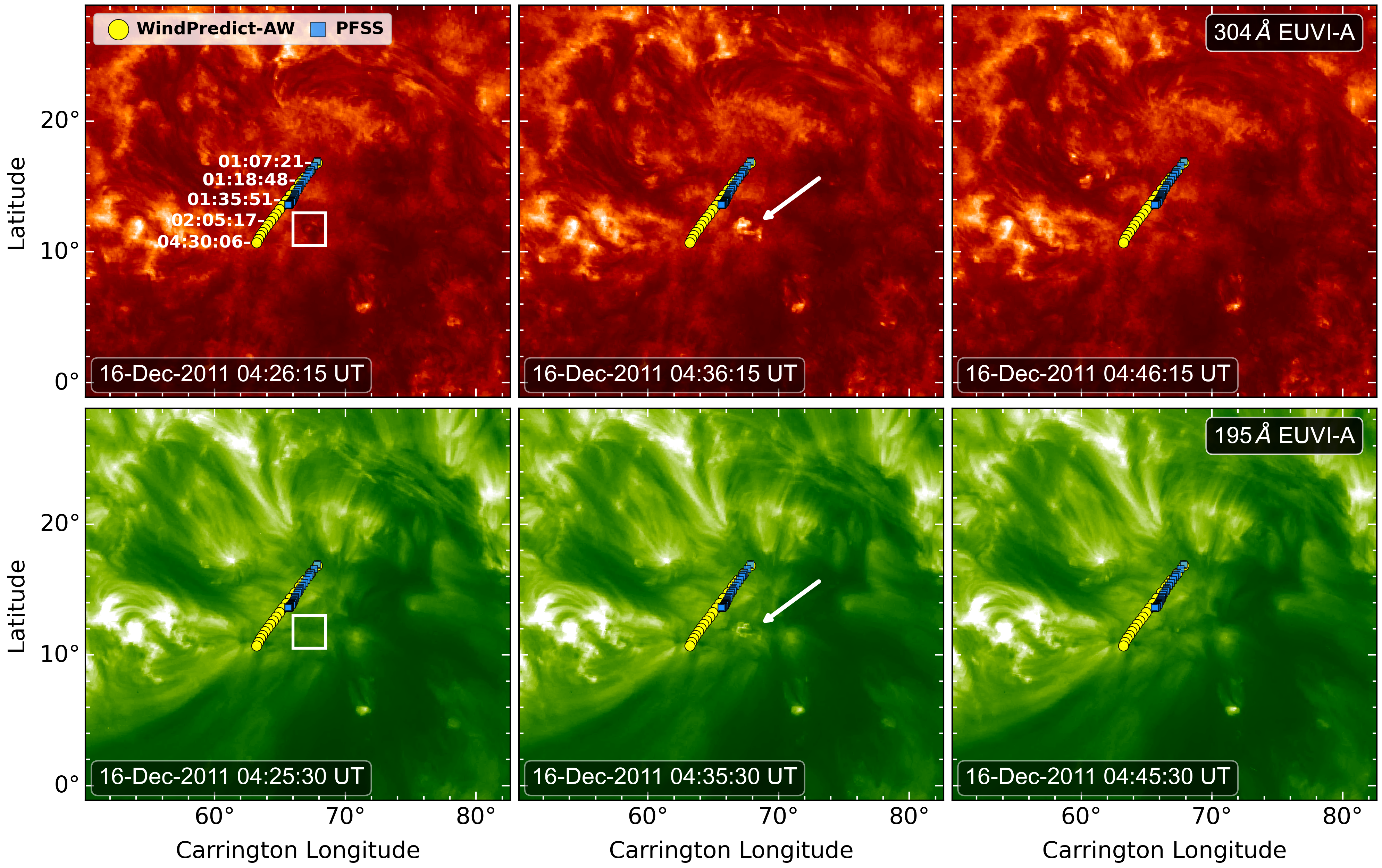}
    \caption{Comparison of the footpoint locations of field lines connecting the comet to the solar surface for different models on December 16, 2011. A SolarSoft synchronic map of the radial magnetic field is used as an input for the PFSS (blue squares) and WindPredict-AW MHD model (yellow dots). Both models use $\ell_{\text{max}} = 50$. The STEREO-A/EUVI data before (left), during (middle), and after (right) the eruption for the $304\AA$ (top row) and $195\AA$ (bottom row) channels are shown. The brightening location is marked by a white arrow in the middle column. In the upper left panel, the time annotations, dated December 16, 2011, correspond to the arrival time at the surface of hypothetical Alfvén waves launched at the site of the comet's trajectory for different field lines, calculated from the MHD model. The white box in the left column is used to calculate the pre-brightening background intensity in Section \ref{section:energetic}.}
    \label{fig:EUVI}
\end{figure*}

\section{Case study: C/2011 W3 (Lovejoy)}\label{section:lovejoy}

\subsection{Orbit and connectivity}\label{section:connectivity}

\par Comet Lovejoy passed through the solar corona with a perihelion distance of roughly 0.2$R_\odot$ from the surface on December 16, 2011. At this distance, well within the sub-Alfvénic solar wind region, magnetic interactions between the comet and the Sun could have occurred. This magnetic connection was probed by the morphology of the cometary tail in EUV emission by \cite{Downs2013} and \cite{Raymond2014}. In the upper right corner of Figure \ref{fig:comet_trajectory}, the tail of the comet during its egress is shown, as observed by the Atmospheric Imaging Assembly (AIA; \citealt{Lemen2012}
) aboard the Solar Dynamics Observatory (SDO; \citealt{Pesnel2012}) with the 171$\AA$ wavelength filter. In the main panel of Figure \ref{fig:comet_trajectory}, we present the magnetic configuration, with field lines connecting the comet to the surface shown in orange and the comet's trajectory\footnote{Horizons Ephemeris Tool: \url{https://ssd.jpl.nasa.gov/horizons/app.html\#/}.} shown in cyan. The coronal magnetic field is reconstructed with the MHD model WindPredict-AW \citep{Reville2020, Reville2022, Parenti2022}, which leverages the use of a transition region as described in \citet{2021sf2a.conf..230R}, with a temperature at the base of the domain of $4\times10^4$ K. The model is set such that the input Poynting luminosity through the bottom boundary is on average $3.3 \times 10^{22}$ W. We refer the reader to \cite{Reville2020} and \cite{Parenti2022} for details on the WindPredict-AW model. The input photospheric boundary condition is given by a synchronic map made available through SolarSoft \citep[hereafter SSW;][]{SSW2000} from December 16, 2011\footnote{Map ID: Bfield\_20111216\_000400.h5.}, and is displayed on the surface (see Figure \ref{fig:comet_trajectory}). Such a map is generated by evolving the magnetic flux of a full-disk magnetogram from the Helioseismic and Magnetic Imager (HMI; \citealt{Scherrer2012}) instrument with a surface flux transport model (SFT; \citealt{Schrijver2003}). To avoid ringing patterns in the magnetic field map used at the boundary of the model due to the spherical harmonics cutoff $\ell_{\rm max}=50$, we used the method of \cite{2025RASTI...4...30M} to smooth the high $\ell$ coefficients with $\alpha=3\times10^{-6}$.

\par By monitoring the solar activity from three vantage points (STEREO-A, STEREO-B, and SDO) during the perihelion, where we expect the SPMI power to be the highest (see Appendix \ref{appendix:spmi_power}), we examined whether any intensity enhancements appeared to be correlated in space and time with the passage of the comet. We identified one tentative event during the egress, with a brightening observed around 04:35 UT on December 16 by the Extreme Ultraviolet Imager (EUVI) \citep{EUVI2024} aboard STEREO-A in the $304\AA$ and $195\AA$ channels. The heliographic Stonyhurst coordinates of the brightening are 101$^{\circ}$ in longitude and 12$^{\circ}$ in latitude, which means they are not visible from Earth. The EUVI image of this brightening in the $304\AA$ filter is shown in Figure \ref{fig:comet_trajectory} at the bottom right. In this panel, the white field lines, whose yellow footpoints fall within $5^{\circ}$ of the brightening location, connect the comet orbit to the surface. In the rest of this paper, we focus on this particular event. 

\par In Figure \ref{fig:EUVI}, we show the EUVI-A data before (left), during (middle), and after (right) the brightening indicated by the white arrow for the $304\AA$ (top row) and $195\AA$ (bottom row) channels. Due to the limited time resolution of the observations, the brightening is only seen in one frame in $304\AA$ (10 minutes cadence) and one frame in $195\AA$ (5 minutes cadence). The yellow dots correspond to the same footpoints shown in the bottom-right panel of Figure \ref{fig:comet_trajectory}, whose field lines were traced from the WindPredict-AW model. We also verified the footpoint locations derived from a simpler PFSS extrapolation (blue squares), using a source surface height of $R_{ss} = 2.5\,R_{\odot}$ and $\ell_{\text{max}} = 50$. Both models use the same SSW input synchronic map. We see a good agreement between them, even though the WindPredict-AW model generates connected field lines that reach lower latitudes. The field line footpoints fall close to the position of the brightening event. As the MHD model agrees well with the PFSS predictions, we used the computationally less expensive PFSS model to explore the sensitivity of the connectivity to different magnetic synchronic maps and PFSS parameters ($\ell_{\text{max}}$ and $R_{ss}$). This is discussed further in Appendix \ref{appendix}, where we show that our results remain robust when varying $R_{ss}$, $\ell_{\text{max}}$ , and the synchronic map date. In the remainder of this analysis, we make use of the WindPredict-AW MHD model to assess the energetics of the hypothetical interaction.

\par We investigate whether the comet could have deposited enough energy at the surface to account for the brightening by examining the spatial and temporal correspondence between its trajectory and the eruption. For the temporal aspect, we assess whether perturbations, modeled as hypothetical Alfvén waves, could propagate fast enough in the solar wind from the comet’s trajectory to the corona and transition region to trigger the eruption. As shown in Figure \ref{fig:mach_orbit}, the WindPredict-AW MHD model indicates that the comet traversed a region where the solar wind was sub-Alfvénic when connected close to the brightening region, with a wind Alfvénic Mach number between 0.16 and 0.78 in the region delimited by the red band. In such a regime, Alfvén waves generated at the comet’s location can propagate upstream, back toward the Sun. Using the local Alfvén speed derived throughout the three-dimensional domain, we computed the travel time of an Alfvén wave from any point along the cometary orbit to a specified height in the solar atmosphere. The travel time was obtained by integrating along selected magnetic field lines while accounting for the background solar wind velocity. In the upper-left panel of Figure \ref{fig:EUVI}, the time annotations indicate the corresponding arrival times of these Alfvén waves in the solar atmosphere for a few field lines. Note that the travel time could only be computed for the WindPredict-AW model and not the PFSS model, since the Alfvén velocity along field lines is necessary. The travel time varies substantially between field lines: the earliest footpoints have associated travel times of only a few minutes, whereas it is on the order of a few hundred minutes for the latest footpoints (see the bottom panel of Figure \ref{fig:distribution}). This spread is caused by the rapid expansion of the magnetic structure connected to the comet as it moves away from the Sun during egress (Figure \ref{fig:comet_trajectory}), which modifies both the Alfvén speed and the propagation distance (varying between roughly one and three solar radii between the first and last footpoints). One field line shows an associated arrival time at 04:30 UT, only minutes before the STEREO-A/EUVI exposure around 04:35 UT on December 16 that observed the brightening. This may have deposited energy near a preexisting magnetic structure, seen as a circular dark feature in the upper left panel of Figure~\ref{fig:EUVI} inside the white box, and thereby triggering a brightening.

\subsection{Energetics}\label{section:energetic}

Once we identified a candidate event of magnetic interaction between the comet and the Sun, we estimated the available energy that could be deposited at the surface to compare with the observations of the brightening. We used the power scaling law of \cite{Paul2026} derived from numerical simulations in the context of SPMIs. Within this framework, the energy flux is transported by perturbations produced when the obstacle and the magnetic field lines connecting it to the host star obstruct the solar wind plasma. In this case, the obstacle is the nucleus of the comet and the ionized envelope around it that extends as a tail. Based on the multi-fluid MHD model of \cite{Jia2014}, the estimated electron density in the cometary tail is of the order of $n_e \approx 10^9-10^{10}$ cm$^{-3}$; this corresponds to an ionization fraction close to unity, which is consistent with the observations reported in \cite{Raymond2014}. Even though the MHD model considers a steady state, the results are still representative as the timescale for ionization, dominated by electron collisions, is below 1 second for sungrazing comets \citep{Jones2018}, which is much quicker than the orbital time inside the corona. Moreover, the MHD simulation of \cite{Jia2014} clearly shows that the comet is an obstacle to the magnetized solar wind, as the field lines drape around the nucleus and the tail. On top of that, \cite{Raymond2014} showed that oxygen ions from the comet could lose up to 40\% of their kinetic energy in the form of waves during perihelion, which implies that magnetic perturbations are possible. The perturbations can travel as Alfvén waves back toward the Sun due to the sub-Alfvénic velocity of the solar wind at the height of the comet during its perihelion \citep[e.g.,][]{Bandyopadhyay2022}. The steady state interaction of a conductive obstacle in a magnetized medium can be estimated in the framework of Alfvén wings \cite{Neubauer1998}. It is not guaranteed that fast magnetic comets can sustain such wings. Nevertheless, the power carried through Alfvén wings provides a useful upper limit for this interaction and this power can be estimated by 
\begin{equation}\label{eqn:SPMI}
    S_{\text{PS26}} = 0.857 \times\left(\frac{\pi \text{R}^2 \,\text{B}_w^2 \, v_{\text{rel}}}{\mu_0}\right)\left(\frac{\text{B}_{\text{obst}}}{1\textrm{T}}\right)^{0.5}\,\left(\frac{1\text{T}}{\text{B}_w}\right)^{0.85}\,\,\,[\text{Watts}],
\end{equation}
\\
where $R$ is the size of the obstacle (roughly 1 Mm for the width of the cometary tail along the orbital direction as modeled by \citealt{Jia2014}), $\text{B}_w$ is the magnetic field of the solar wind at the location of the comet, $v_{\text{rel}}$ is the relative velocity between the comet and the solar wind, $\text{B}_{\text{obst}}$ is the magnetic field of the obstacle, and 1T corresponds to units in Tesla. Although the comet does not posses an intrinsic magnetosphere generated by dynamo processes, the electrically conductive ionized tail can still perturb the flow of the solar wind. In this induced magnetosphere scenario \citep[e.g.,][]{Luhmann2004}, we can approximate $\text{B}_{\text{obst}}\approx \text{B}_w$, which assumes a magnetosphere of the same size as the obstacle itself. It is worth mentioning that analytical models also exist for the power estimate by SPMIs. In the case of the \cite{Saur2013} model, an upper limit for the power, assumed to be channeled by a Poynting flux along Alfvén wings, is given by
\begin{equation}\label{eqn:saur}
    S_{\text{S13 (Max)}} = \frac{6\pi\,R^2\,v_{\text{A}}\,(\text{M}_{\text{A}}\text{B}_w)^2}{\mu_0}\,\,\,[\text{Watts}],
\end{equation}
where $v_{\text{A}}$ is the Alfvén velocity and $\text{M}_{\text{A}} = v_{\text{rel}}/v_{\text{A}}$ is the Alfvénic Mach number of the moving obstacle. This model assumes that in the reference frame of the moving obstacle, the flow of the solar wind around the obstacle is sub-Alfvénic ($M_A \rightarrow 0$) and gives values close to the original idea proposed by \citet{Zarka2007}, up to a factor of $2M_A$ \citep{Saur2013}. However, even though the comet was orbiting in a region of sub-Alfvénic solar wind for most of its perihelion (see Figure \ref{fig:mach_orbit}), the comet itself was moving at a super-Alfvénic velocity according to the WindPredict-AW model. The only exception is between approximately 00:52 UT and 02:00 UT, when the comet Alfvén Mach number drops below unity, but still does not approach zero. Therefore, the condition $M_A \rightarrow 0$ for the moving comet is not satisfied for the entirety of the section of the orbit considered in our analysis, which makes this model not ideal for our case study. As for the "stretch-and-break" model of \cite{Lanza2013}, the SPMI power is given by
\begin{equation}\label{eqn:lanza}
    S_{\text{L13}} = \frac{2\pi f\,\text{R}^2\,\text{B}_{\text{obst}}^2\,v_{\text{rel}}}{\mu_0}\,\,\,[\text{Watts}],
\end{equation}
where $f$ is the fraction of the obstacle's surface magnetically connected to the Sun. In our assumption of an induced magnetosphere ($\text{B}_{\text{obst}}\approx \text{B}_w$), this parameter has a value of unity.

\par We estimated the power available along the field line whose Alfvén wave arrival time (04:30 UT) is just before the brightening. The whole power distribution for all field lines near the brightening is shown in Figure \ref{fig:distribution}. We used Equations \ref{eqn:SPMI}-\ref{eqn:lanza} to get a range of calculated SPMI powers. At the comet location along this field line, the relative velocity and magnetic field strength are $v_{\text{rel}} \simeq 403\,\text{km}$ s$^{-1}$ and $\text{B}_w \simeq 5.3\times10^{-6}\,\text{T,}$ respectively, giving powers $S_\text{PS26} \simeq 4\times10^{14}\,\text{W}$, $S_\text{S13 (Max)} \simeq 6\times10^{13}\,\text{W}$, and $S_\text{L13} \simeq 10^{13}\,\text{W}$. The discrepancy between the models is further discussed in Section \ref{section:discussion}. We use the power value $S_\text{PS26}$ for the rest of the analysis as it is derived from 3D MHD numerical simulations, and also gives the largest power. Note that we do not consider the decay or the reflection of Alfvén waves at the transition region as this would lead part of the waves being transported away from the Sun before the energy deposition \citep[e.g.,][]{Paul+2025}. As a result, our estimate only provides an upper limit for the transferred power. 

\par In the bottom-left panel of Figure \ref{fig:EUVI}, the white box outlines the region used to estimate the pre-brightening background intensity. We subtracted this background level from the intensity during the brightening and then summed over all pixels with a signal over 25 DN s$^{-1}$. Varying this threshold value does not significantly affect the resulting total intensity. We find a value in the EUVI 195$\AA$ channel of $I_{195,\text{obs}} \simeq 2500~\text{DN}~\text{s}^{-1}$. We used this value for a comparison with the expected intensity if all the SPMI power $S_{\text{PS26}}$ deposited on the area is converted into radiation. We used a first approximation where all the emission is covered by the 195$\AA$ wavelength band. We are aware that this is possibly not the case, thus we provide an upper limit on the expected emission in this wavelength range. Since the EUVI instrument is a non-monochromatic imager, we cannot isolate the energy of the photon falling on the detector. As a first approximation, we thus assumed that all photons are emitted at a single wavelength $\lambda_{\text{peak}} = 198\AA$, corresponding to the peak of the EUVI 195$\AA$ effective area. We calculated this function using the \texttt{euv\_parms.pro} function available in SSW, and setting the inputs as obtained from the observation at the time of the brightening. The emission, assumed to be isotropic, originates from a brightening region with a projected area of $A_{\text{brightening}} \simeq 230~\text{Mm}^2$, as measured from Figure \ref{fig:EUVI}. Under these assumptions, the radiative power at the emission location $S_{\text{rad}}$ can be written as
\begin{equation}\label{eqn:intensity}
    S_{\text{rad}} = \frac{4\pi~A_{\text{brightening}}~I_{195,\text{obs}}}{R_{195}(\lambda_{\text{peak}})}\times \frac{hc}{\lambda_{\text{peak}}}\, ,
\end{equation}
where $R_{195}(\lambda_{\text{peak}}) \simeq 3.94 \times 10^{-12}~\text{DN}~\text{cm}^2~\text{sr}~\text{ph}^{-1}$ is the peak instrumental response function. The calculated radiative power is $S_{\text{rad}} \simeq 2\times10^{17}~\text{W}$. We can also estimate the modeled intensity corresponding to the SPMI power value of $S_{\text{PS26}} \simeq 4\times10^{14}~\text{W}$ by inverting Equation \ref{eqn:intensity}. This gives an intensity of $I_{\mathrm{SPMI}} \simeq 6~\mathrm{DN}~\mathrm{s}^{-1}$, which is just above the estimated photon noise \citep{Gonzalez2016} and represents only a few percent of the averaged EUVI 195$\AA$ intensity measured at the brightening location prior to the onset of the brightening (roughly 190 DN s$^{-1}$). Therefore, we can conclude here that the potential magnetic interaction due to the comet's passage cannot provide enough power to be the energy source of the observed brightening. We discuss the alternative scenario of the perturbation of an existing magnetic structure in the next section. 

\section{Discussion and future comets}\label{section:discussion}

\subsection{Hotspot formation versus flare triggering due to SPMIs}

\par The 195$\AA$ radiative power estimated in Section \ref{section:energetic} is significantly larger than the SPMI power predicted by the \cite{Paul2026} model. We note that emission in the 304$\AA$ band is not included in this estimate, since the plasma is optically thick for this emission line, and absorption effects are non-negligible. We would need to solve a radiative transfer problem to simulate the emission in this band, which is outside of the scope of this paper. Our approach represents a significant simplification of the radiative energy budget. As a result, the hotspot scenario as the cause for the brightening is not supported by our observations. The only plausible influence of the comet would therefore be through perturbation of a pre-existing magnetic structure, potentially triggering its eruption \citep{Ilin2025}. If we consider the event to be a regular flare, based on its size, duration, and intensity, we estimate it to be equivalent to a GOES B-class flare or lower. Such flares are expected to statistically occur \citep{Veronig2002} during the time window of the comet interaction (a few hours, see Figure \ref{fig:mach_orbit}). Therefore, to rule out this event as a regular flare, one would need a more accurate estimate of the connectivity and a higher temporal cadence of the multiwavelength light curve. In addition, at the time of comet Lovejoy’s perihelion, the Sun was also quite active, which makes it difficult to unambiguously identify magnetic interactions given the limited observations of this event.

\par Magnetic interactions between a comet and the Sun could, in principle, produce radio emission \citep[similar to SPMIs, see][]{Zarka1998, Zarka2018, Callingham2025}. Such observations are not addressed in this study, as the brightening event of interest was not visible from Earth. Although our results do not support the hotspot scenario, these magnetic interactions could still accelerate energetic electrons, potentially generating radio bursts or auroral emission above regions of strong magnetic fields, as reported by \cite{Yu2024} who observed radio emission above a sunspot due to precipitating electrons accelerated by nearby flares. This is an interesting line of inquiry for future sungrazing comets.

\par As shown in Figure \ref{fig:power_orbit}, the predicted SPMI power remains comparable during both the ingress and egress phases. However, it reaches a maximum of approximately 10$^{17}$ W near perihelion, where the comet’s velocity and the solar wind magnetic field strength are both at their highest. This peak value is consistent with the estimated radiative power derived in Section \ref{section:energetic}. If we consider this maximum SPMI power value, the corresponding modeled intensity is $I_{\mathrm{SPMI,~\text{max}}} \simeq 1300~\mathrm{DN}~\mathrm{s}^{-1}$, which would be readily detectable by the EUVI instrument, well above the estimated noise level of a few DN s$^{-1}$ \citep{Gonzalez2016}. Unfortunately, at perihelion, the footpoints of the magnetic field lines connecting the comet to the Sun were located near the solar limb from the perspectives of both STEREO-A and STEREO-B. For the few visible footpoints around perihelion, no brightenings were observed. This means that in the energy deposition scenario, a maximum of 0.1\% of the SPMI power would have to be deposited to leave a signal below the noise level of EUVI and to explain its non-detection. Along those footpoints, no pre-existing magnetic structures susceptible to eruptions were observed, which implies that no flare triggering scenario could take place. As a result, we were unable to identify any brightenings or flaring activity associated with the comet’s passage during perihelion. 

\par A clear detection of comet-Sun magnetic interactions as observed from Earth's point of view, with SDO/AIA for instance, would be significant in improving our understanding of energy channeling and deposition due to SPMIs. Unlike observations of exoplanetary systems, with a sufficiently high temporal cadence it could be possible to see the heating sequence at different heights in the atmosphere as seen in different EUV channels. This would help discriminate between the two proposed scenarios: an intensity increase appearing first at higher altitudes would favor the energy deposition (hotspot) scenario, whereas an increase originating at lower altitudes could be consistent with either a hotspot or a flare-triggering mechanism. The typical flare profile as a function of altitude for different channels is well known. In the energy deposition scenario, we would expect to see a significantly different pattern with a clear phase delay from high to low altitudes. Observations of the magnetic field with SDO/HMI or Solar Orbiter/PHI, for example, would also provide crucial information about the magnetic connectivity for the region of interest. 

\subsection{Applicability of SPMI models to sungrazing comets}

\par
The strongest SPMIs occur when the obstacle perturbing the solar wind has a magnetosphere or is electrically conductive. For comets, the ionized envelope around the nucleus, which extends as a tail, can act as this conductive obstacle. For comet Lovejoy, the ionization fraction is inferred to be close to unity \citep{Jia2014, Raymond2014}, due to the rapid ionization of sublimated gases in the hot coronal environment \citep[e.g.,][]{Downs2013}. Comets passing farther from the Sun encounter cooler temperatures and lower coronal densities, potentially leading to lower ionization fractions and a reduced ability to perturb magnetic field lines and drive SPMIs (as long as they still orbit within a region of sub-Alfvénic solar wind). In this case, the assumption that $\text{B}_{\text{obst}}\approx \text{B}_w$ could be invalid. Moreover, the transmitted power in Equation~\ref{eqn:SPMI} also depends on the size of the obstacle, which introduces uncertainty through assumptions about the tail geometry. Here we assume a simple cylindrical geometry following \citet{BryansPesnell2012}, with a circular cross-section of roughly 1 Mm for the envelope. However, observations of comet Lovejoy show that its tail morphology was shaped by the local magnetic field, with striations \citep{Raymond2014} and an offset between the direction of motion and the plasma tail \cite{Downs2013}, which are effects not captured by this simplified model. 

\par As shown in Section \ref{section:energetic}, the SPMI power that could be transferred by the comet varies by up to two orders of magnitude depending on the model used \citep{Lanza2013,Saur2013,Paul2026}. Such a discrepancy is not unexpected given the results shown in Figure 4 of \cite{Paul2026}. The different models predict different scaling laws that diverge as the magnetic field strength changes. In the calculation performed in Section \ref{section:energetic}, the solar wind magnetic field value of $\text{B}_w = 3.4\times10^{-6}~\text{T}$ lies well below the range explored in Figure 4 of \cite{Paul2026}. Extrapolating the scaling laws to this low-field regime, we therefore expect significant divergence between the models. In particular, the \cite{Lanza2013} model, which has the steepest scaling, would probably give the lowest SPMI power, whereas the \cite{Paul2026} model, with the shallowest scaling, would predict the highest power, which is consistent with our results. This trend is also observed for the whole power distribution when considering all magnetic field lines of Figure \ref{fig:EUVI}, as further discussed in Appendix \ref{appendix_powers}. 

\subsection{Other possible candidates for comet-induced SPMIs}

\par Several sungrazing comets have passed close to the Sun over the past decade or so (see \citealt{Jones2018} for a list). In addition to comet Lovejoy, we examined two such events, C/2011 N3 (SOHO) and C/2012 S1 (ISON), but found no evidence of intensity enhancements induced by magnetic interactions. In the case of C/2011 N3 (SOHO), the nucleus disintegrated within the solar atmosphere and did not survive perihelion \citep{Schrijver2012}, which resulted in a shorter interaction time with the coronal magnetic field compared to comet Lovejoy. For C/2012 S1 (ISON), no EUV emission was detected by AIA. Although its perihelion distance was only moderately larger than that of Lovejoy (by a factor of $\sim$8.8, but still close enough to be in a region of sub-Alfvénic solar wind), \citet{BryansPesnell2016} attributed the lack of emission to its smaller nucleus, which was estimated to be about four times smaller than Lovejoy. This reduced nucleus size would result in a smaller plasma tail and, consequently, weaker power transmission through SPMIs. Using the Minor Planet Center database of known comets from the International Astronomical Union\footnote{\url{https://minorplanetcenter.net/data}}, we also searched for future candidate comets. Over the next decade, we only identified one such object: comet 322P/SOHO, which will reach a perihelion distance of 10.9 solar radii on August 11, 2027. Due to its larger perihelion distance, this object is classified as a sunskirting comet rather than a sungrazer \citep{Jones2018}. Other candidates might also be discovered in the upcoming years.

\section{Conclusion}\label{section:conclusion}

The passage of comet Lovejoy through the solar corona provided a rare opportunity to investigate star–planet magnetic interaction (SPMI) processes within our own solar system. Using the reconstructed magnetic connectivity between the comet and the Sun, we estimated the energy that could be transferred to the solar surface via Alfvén waves. Focusing on a brightening event that is spatially and temporally consistent with the comet’s passage, we find that the most plausible scenario is not direct energy deposition, but rather a magnetic perturbation capable of triggering an eruption in an already stressed configuration. Further observations of future sungrazing comets will be essential to test this scenario and better constrain the physics of comet–Sun magnetic interactions.

\begin{acknowledgements}
    This work utilizes data produced collaboratively between AFRL/ADAPT and NSO/NISP. This work used data provided by the MEDOC data and operations centre (CNES / CNRS / Univ. Paris-Saclay), \url{http://medoc.ias.u-psud.fr/}. This research has made use of data and/or services provided by the International Astronomical Union's Minor Planet Center. This work has been supported by the  European Research Council through the Consolidator Grant number 101125367 (``ExoMagnets'', ERC-2023-CoG), the French Agence Nationale de la Recherche (ANR) project STORMGENESIS \#ANR-22-CE31-0013-01, the European Research Council through the Synergy Grant number 810218 (``The Whole Sun'', ERC-2018-SyG),  the Centre National d’Etudes Spatiales (CNES) Solar Orbiter project, the Institut National des Sciences de l’Univers (INSU) via the Action Thématique Soleil-Terre (ATST), and Programme for Supercomputing, and by computing HPC and storage resources by GENCI thanks to the grant 2025-A0180410133. We thank F. Auchère for useful discussions on the STEREO-A/EUVI  instrument.
\end{acknowledgements}  

\bibliographystyle{aa}
\bibliography{BIB}

\begin{appendix}

    \section{Distribution of SPMI power near the brightening}\label{appendix_powers}

    \par
\begin{figure}[h!]
            \centering
            \includegraphics[width=0.75\columnwidth]{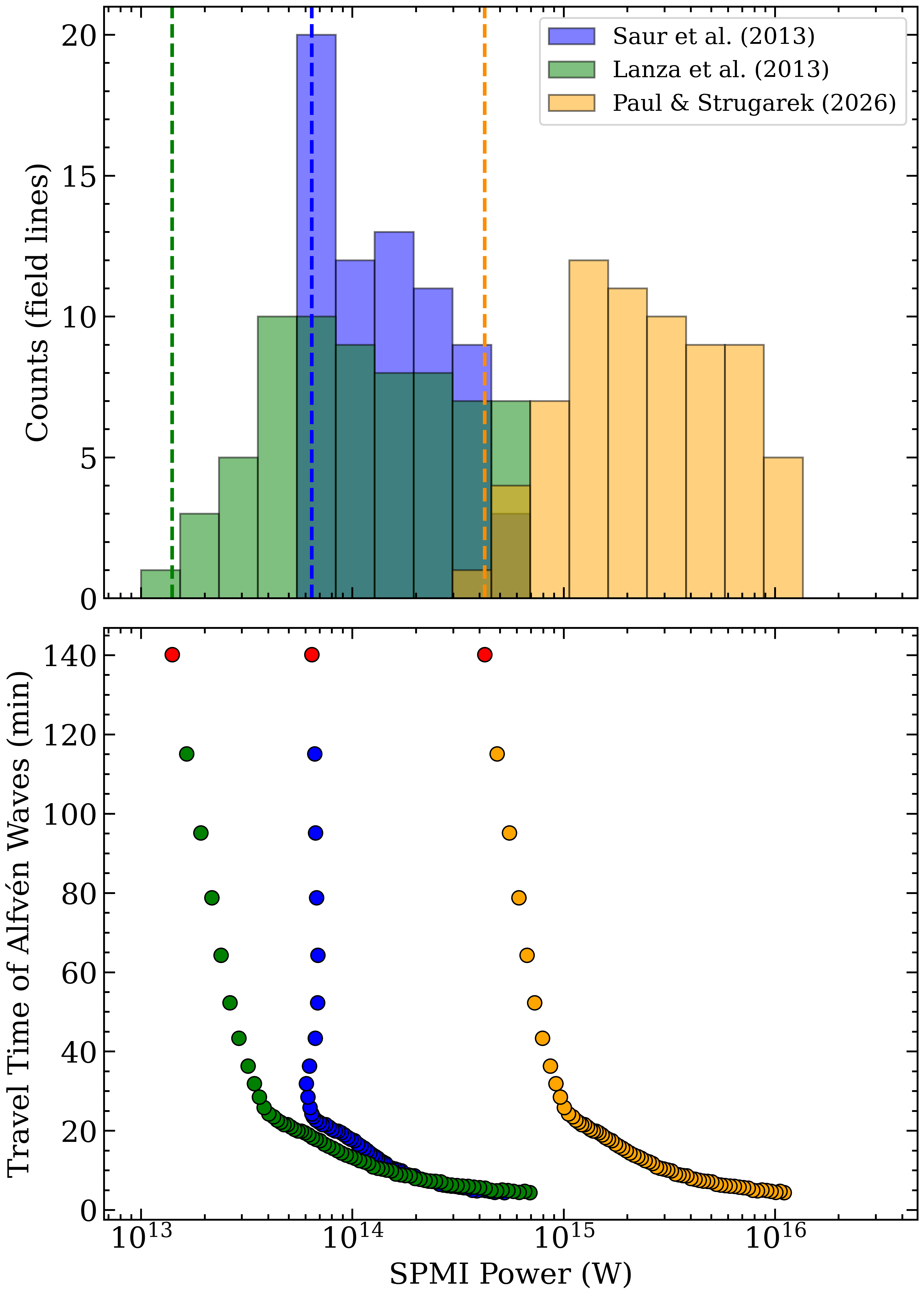}
            \caption{Top: Distribution of SPMI power calculated with the \cite{Saur2013} (blue), \cite{Lanza2013} (green), and \cite{Paul2026} (orange) models for all field lines in Figure \ref{fig:EUVI} whose footpoints lie within $5^{\circ}$ of the brightening. The field lines are generated from the minute cadence of the comet's trajectory. The single-field-line values from Section \ref{section:energetic} are indicated by vertical dashed lines. Bottom: Propagation time of hypothetical Alfvén waves for each magnetic field line as a function of the associated SPMI power. The red points correspond to the powers labeled with vertical dashed lines in the upper panel.}
            \label{fig:distribution}
            \end{figure}

 In Section \ref{section:energetic}, we computed the SPMI power for a single magnetic field line whose hypothetical Alfvén waves reach the Sun just before the brightening. The top panel of Figure \ref{fig:distribution} shows the distribution of SPMI powers obtained with the \cite{Saur2013} (blue), \cite{Lanza2013} (green), and \cite{Paul2026} (orange) models for all field lines anchored within $5^{\circ}$ of the brightening (Figure \ref{fig:EUVI}) with an Alfvén wave arrival time before the event. The values from Section \ref{section:energetic} are indicated by vertical dashed lines and lie at the low end of each distribution. This reflects the fact that, at the time of the brightening, the comet is moving away from the Sun and connected to a rapidly expanding magnetic structure, leading to decreasing $v_{\mathrm{rel}}$ and $B_w$ along successive field lines and thus lower SPMI powers (Equation \ref{eqn:SPMI}). For the \cite{Paul2026} model, the distribution extends up to roughly above $10^{16}~\mathrm{W}$, approaching the estimated radiative power. Note that $10^{16}~\mathrm{W}$ is the maximal SPMI power for field lines near the brightening, not when considering the whole perihelion orbit (discussed in Appendix \ref{appendix:spmi_power}). 
        
    The bottom panel shows the Alfvén wave travel time as a function of SPMI power. Higher SPMI powers, primarily associated with larger $B_w$, as $v_{\mathrm{rel}}$ 
    does not vary significantly along this part of the orbit, correspond to shorter travel times, consistent with faster Alfvén propagation in stronger magnetic fields.

    \section{Properties along the comet's orbit near perihelion}\label{appendix:orbit}

    \subsection{Solar wind Alfvénic Mach number}

\begin{figure}[h!]
            \includegraphics[width=0.95\columnwidth]{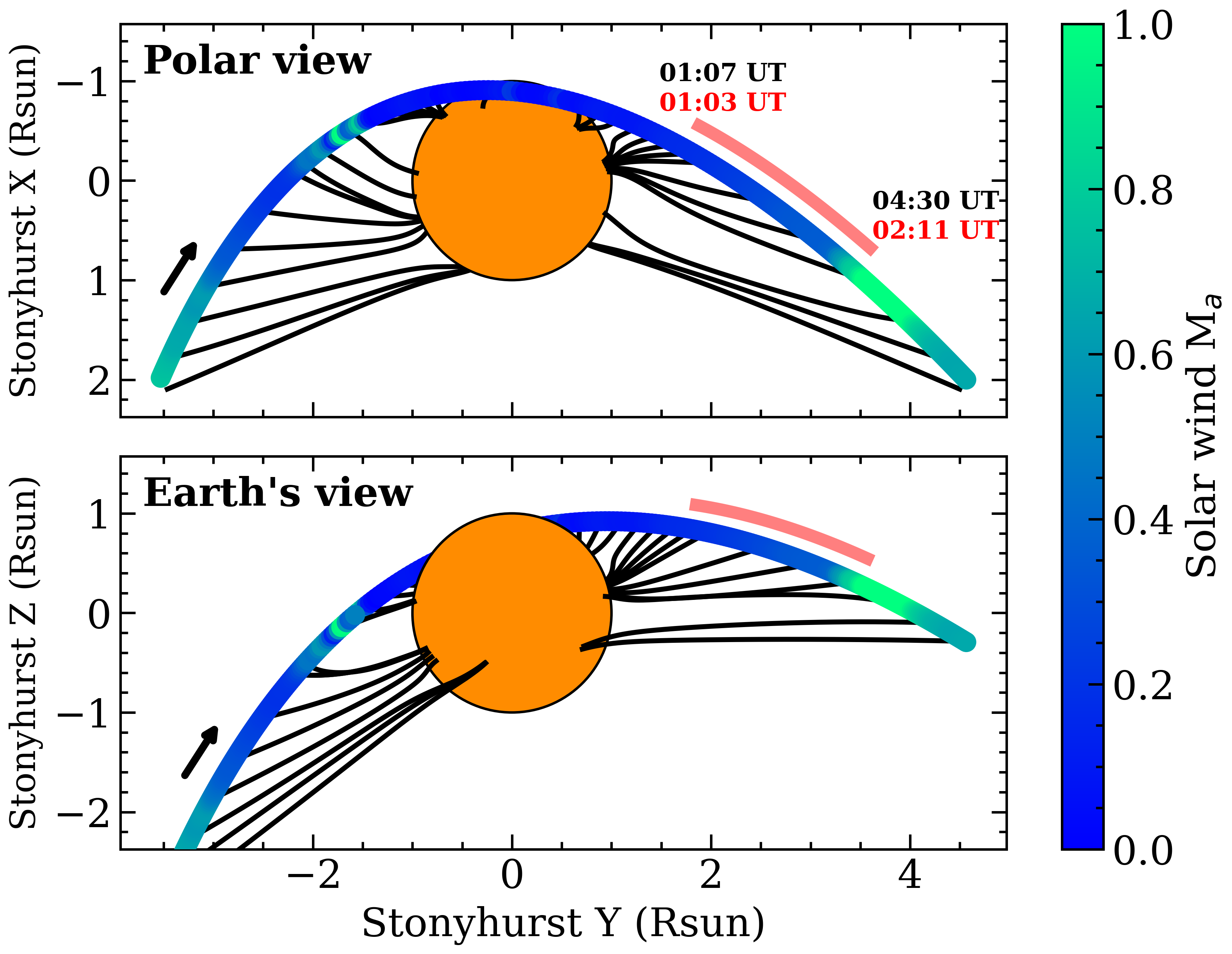}
        \caption{Alfvénic Mach number of the solar wind computed from the WindPredict-AW model along the orbit of comet Lovejoy nears its perihelion. The orbit is viewed from a polar view (top) and from Earth's view (bottom) in Heliographic Stonyhurst coordinates expressed in cartesian form. The orbital direction is indicated by the black arrow and the Sun as the orange disk. The magnetic field lines connecting the comet's trajectory to the Sun are shown as black lines. The red segment delimits the portion of the orbit connected to the Sun near the brightening. The red timestamps correspond to the orbital time of the comet, whereas the black timestamps correspond to the arrival time at the surface of Alfvén waves launched from that position in the orbit.}
        \label{fig:mach_orbit}
    \end{figure}

     Figure \ref{fig:mach_orbit} shows the Alfvénic Mach number of the solar wind computed from the WindPredict-AW model along the orbit of comet Lovejoy near perihelion. Over most of the orbit, the solar wind remains sub-Alfvénic, except during the end of the egress, where the flow becomes super-Alfvénic. In this region, the rapid expansion of the magnetic structure reduces the magnetic field strength, which in turn lowers the local Alfvén speed. As a result, the solar wind speed exceeds the Alfvén speed, even if the radial distance from the Sun is only a few solar radii. Importantly, the segment of the orbit magnetically connecting the comet to the Sun near the location of the brightening (indicated by the red band) does not lie within this region of super-Alfvénic solar wind. Therefore, Alfvén waves generated at the comet’s position could still propagate back toward the solar surface.

    \subsection{SPMI power}\label{appendix:spmi_power}

\begin{figure}[h!]
                \includegraphics[width=0.9\columnwidth]{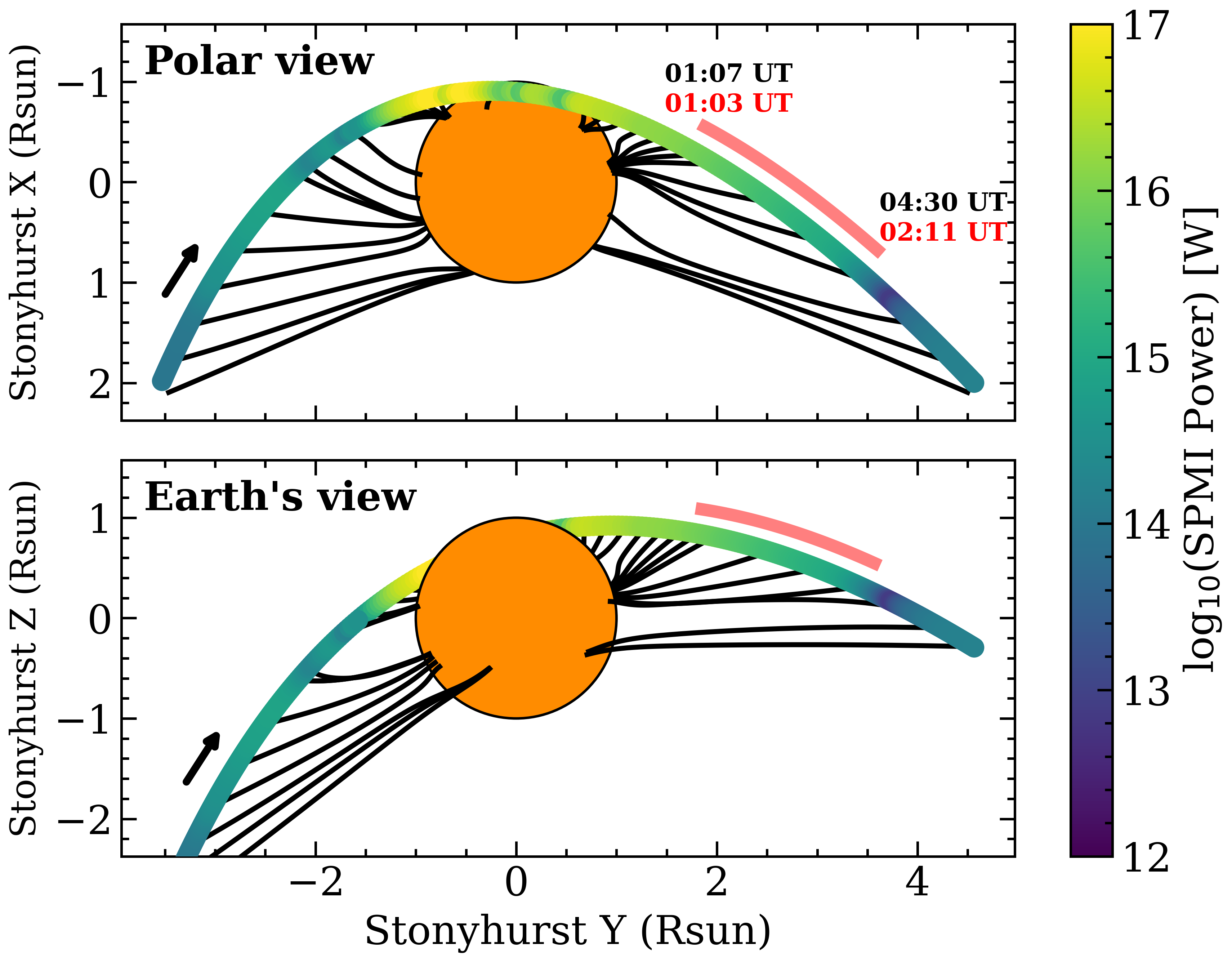}
            \caption{Same as in Figure \ref{fig:mach_orbit}, but for the SPMI power along the orbit.}
            \label{fig:power_orbit}
        \end{figure}

       Figure \ref{fig:power_orbit} is similar to Figure \ref{fig:mach_orbit}, but shows the SPMI power computed with the \cite{Paul2026} model (Equation \ref{eqn:SPMI}). As expected, the SPMI power peaks near perihelion (roughly $10^{17}~\text{W}$), where both the comet’s velocity and the solar wind magnetic field strength are maximal. This value is slightly over the upper end of the power distribution of field lines near the brightening (Figure \ref{fig:distribution}) and is consistent with the estimated radiative power of the brightening calculated in Section \ref{section:energetic}. We note that many magnetic field lines connect to the surface near the brightening region, meaning that even though the SPMI power is not the highest in this section of the orbit, more energy could have been deposited there compared to the region connected to the perihelion.
                
    \section{Influence of model parameters on the connectivity}\label{appendix}

    \subsection{Input magnetic synchronic maps}
\begin{figure}[h!]
        \includegraphics[width=0.85\columnwidth]{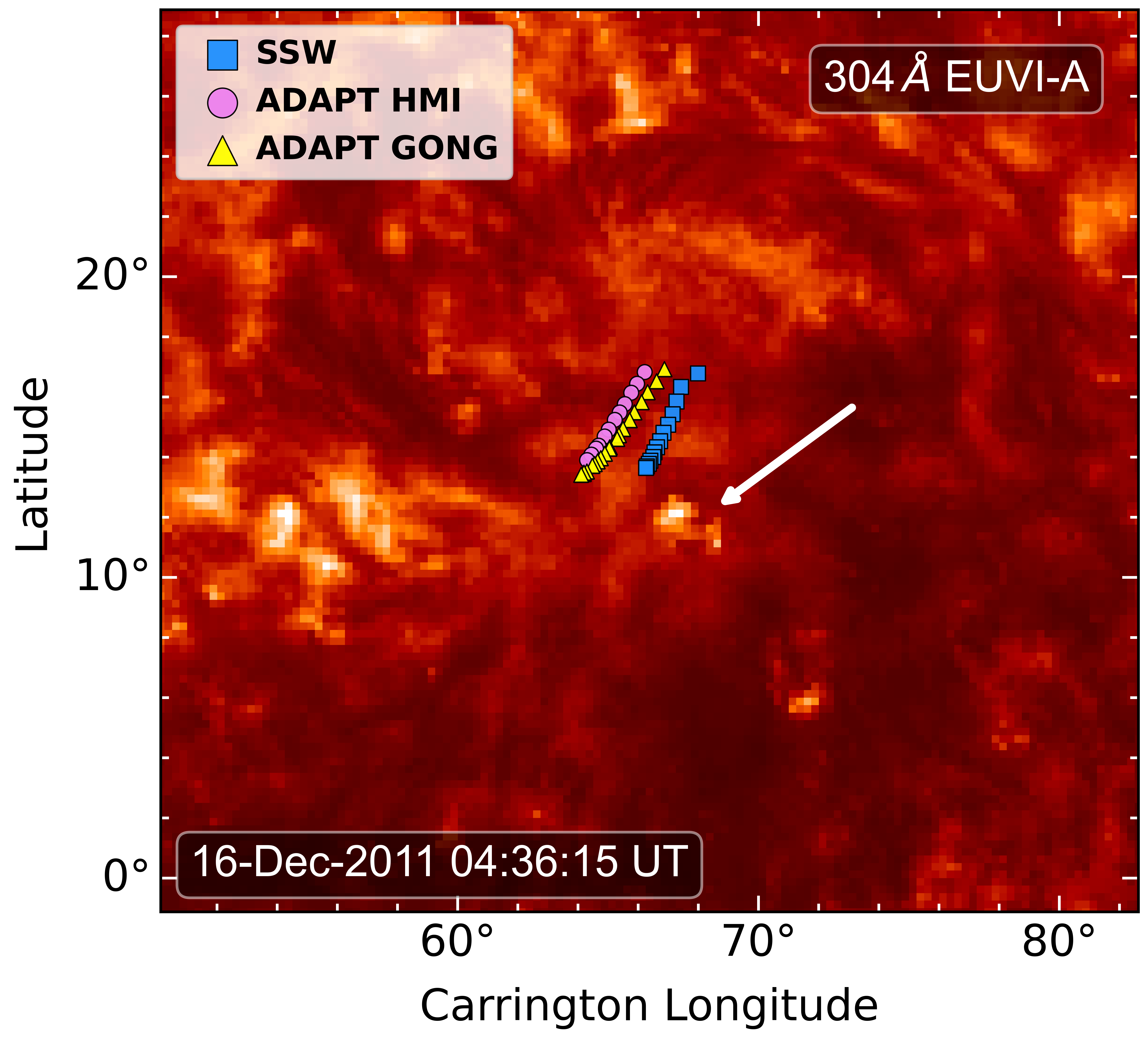}
        \caption{Comparison of footpoint locations for PFSS extrapolations ($R_{ss} = 2.5\,R_{\odot}$ and $\ell_{\text{max}} = 88$) using SSW (blue squares), ADAPT HMI (pink dots) and ADAPT GONG (yellow triangles) synchronic magnetic maps from December 16, 2011, as the photospheric boundary condition. Although the SSW maps allow for a slightly higher resolution ($\ell_{\text{max}} = 94$), we adopt $\ell_{\text{max}} = 88$, i.e., the maximum value permitted by the angular resolution of the ADAPT maps, to ensure a consistent comparison among the three maps. The STEREO-A/EUVI $304\AA$ image of the brightening on December 16 at 04:36 UT is also shown, with the brightening location marked by the white arrow.}
        \label{fig:ADAPT_vs_SSW}
    \end{figure}

     In Figure~\ref{fig:EUVI}, we used the Bfield\_20111216\_000400.h5 SSW map as the photospheric boundary condition for the PFSS extrapolation. We now extend this analysis by performing PFSS extrapolations based on other synchronic magnetic maps, specifically Air Force Data Assimilative Photospheric Flux Transport (ADAPT) maps \citep{Worden2000, Hickmann2015, Arge2010}, using both HMI and Global Oscillation Network Group (GONG; \citealt{Harvey1996}) input magnetograms. Figure~\ref{fig:ADAPT_vs_SSW} compares the resulting connectivity footpoints for the SSW (blue squares), ADAPT HMI (pink dots), and ADAPT GONG (yellow squares) maps on December 16, 2011. The strong agreement among all three cases indicates that the inferred magnetic connectivity is not sensitive to the choice of synchronic map, demonstrating the robustness of our results.

    \subsection{Input PFSS parameters}

    \begin{figure*}[h!]
        \centering
        \includegraphics[width=2\columnwidth]{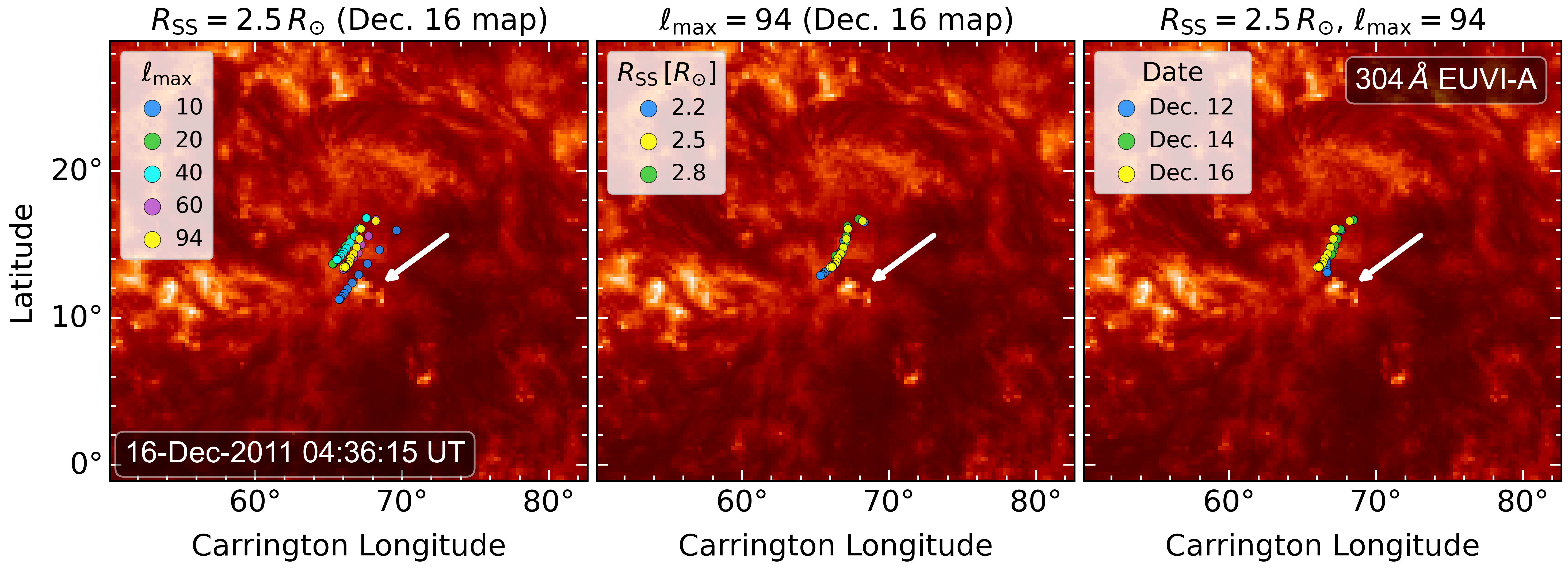}
        \caption{Comparison of footpoint locations for PFSS extrapolations using different values of $\ell_{\text{max}}$ (left), $R_{ss}$ (middle), and SSW synchronic map dates (right). The STEREO-A/EUVI $304\AA$ image of the brightening on December 16 at 04:36 UT is also shown, with the brightening location marked by the white arrow. In the right panel, although the date of the input synchronic map differs from the time of the event, each field line footpoint is projected at the correct Carrington coordinates during the brightening.}
        \label{fig:pfss_comparison}
    \end{figure*}

    \par We now evaluate the effect of using different input parameters on the PFSS extrapolations to test the robustness of the inferred footpoint locations and comet connectivity. In Figure \ref{fig:pfss_comparison}, we compare the footpoint locations for different values of $\ell_{\text{max}}$ (left), $R_{ss}$ (middle), and SSW synchronic map dates (right). We use SSW maps as we have seen in Figure \ref{fig:ADAPT_vs_SSW} that the choice of map source does not make a significant difference in our results. In the left panel, the footpoint locations are consistent with each other, except for the coarser angular resolution with $\ell_{\text{max}} = 10$. Note that $\ell_{\text{max}} = 94$ is the highest possible value based on the resolution of the input SSW synchronic map. In the middle panel, the $R_{ss}$ value does not affect the connectivity.

    \par In the right panel, we test different SSW synchronic magnetic map dates because the SFT model used to generate them assimilates observational data only within $\pm\,60^{\circ}$ from the central meridian \citep{Barnes2023}. On December 16, the brightening site was located at 101$^{\circ}$ from the central meridian, meaning that the magnetic flux at this location was estimated by the model rather than directly constrained by observations. To assess whether this introduces additional uncertainties, we also use a synchronic map from a time when the brightening location lies within the assimilation window, around December 12. Although the date of the input synchronic map differs from the time of the brightening, each field line footpoint is reprojected at the correct Carrington coordinates during the brightening. We see a good agreement for the footpoint locations between the different input synchronic maps, meaning that the magnetic flux around the brightening region does not evolve a lot between December 12 and 16 and the model uncertainties remain small. These results confirm the robustness of the magnetic connectivity used throughout this paper. 

\end{appendix}

\end{document}